\documentstyle[11pt,newpasp,twoside,epsf]{article}
\reversemarginpar

\begin{document}
\title{Current Status of the Microlensing Surveys}
\author{Bohdan Paczy\'nski}
\affil{Princeton University Observatory, Princeton, NJ 08544, USA}

\begin{abstract}
The ongoing microlensing searches have generated more photometric measurements
of pulsating stars than all previous observing projects combined.  In 
particular, OGLE has made $ \sim 340,000 $ $B$, $V$, and $I$-band measurements 
of $ \sim 1,300 $ Cepheids in the Large Magellanic Clouds accessible over 
Internet.
Microlensing searches contributed to the development of very efficient
image subtraction software which works best in crowded fields.  This
suggests the use of a period -- flux amplitude rather than period -- luminosity 
relation for the Cepheids for distance determination, as the flux amplitude
is directly measurable with the image subtraction, and it is not biased
by crowding.  Future projects will dramatically increase the data rate, 
will provide all-sky coverage and a complete census of variables, including 
pulsating stars, to the ever fainter limits.  Time will show which approach,
a small number of large teams or a large number of small teams, 
will be more productive.
\end{abstract}

\section{Introduction}

Three groups: EROS, MACHO and OGLE, reported the detection of their
first gravitational microlensing events in September of 1993 (Aubourg et al.
1993, Alcock et al. 1993, Udalski et al. 1993).  By now several hundred
microlensing events have been discovered, most of them towards the galactic bulge,
some towards the Magellanic Clouds, and a few in other directions.  The
events are detected in real time, and are reported on the World Wide Web at
the rate of approximately two per week.  A mini-collaboration, DUO, reported 
the detection of 13 microlensing events (Alard \& Guibert, 1997), one of 
them due to a double lens (Alard et al. 1995).  The microlensing projects 
monitor tens of millions of stars, they have discovered over one hundred
thousand variables, and over ten thousand pulsating stars.  Information 
may be found at the Web sites: \\

\noindent {OGLE, Poland:   http://www.astrouw.edu.pl/\~~ftp/ogle }\\
{OGLE, USA: http://www.astro.princeton.edu/\~~ogle }\\
{MACHO: http://wwwmacho.mcmaster.ca/}\\
{EROS: http://www.lal.in2p3.fr/recherche/eros/}\\

\noindent Several new groups joined the search for, as well as the follow-up of the microlensing events.  Their Web sites are:\\

\noindent 
{MEGA: http://www.astro.columbia.edu/\~~arlin/MEGA}\\
{MOA: http://www.phys.vuw.ac.nz/dept/projects/moa/}\\
{AGAPE: http://cdfinfo.in2p3.fr/Experiences/AGAPE/}\\
{PLANET: http://www.astro.rug.nl/\~~planet/}\\

\noindent In addition, several other projects generate a huge volume of the photometric data.  The following are the Web sites I was able to find:\\

\noindent {DIRECT: http://cfa-www.harvard.edu/\~~kstanek/DIRECT/}\\
{ROTSE: http://rotsei.lanl.gov/}\\
{ROTSE: http://umaxp1.physics.lsa.umich.edu/\~~mckay/rsv1/rsv1\_home.htm}\\
{LOTIS: http://hubcap.clemson.edu/\~~ggwilli/LOTIS/}\\
{ASAS: http://www.astrouw.edu.pl/\~~gp/html/asas/asas.html}\\
{Ystar: http://csaweb.yonsei.ac.kr/\~~byun/Ystar/}\\
{STARE: http://www.hao.ucar.edu:80/public/research/stare/stare\_synop.html} \\

\noindent I expect to update the list as new projects come along, and they can be found on my home page at: {http://www.astro.princeton.edu/faculty/bp.html}.

In this paper some results obtained with the data generated by the
microlensing projects are described.  As I am not actively working 
on pulsating stars the choice is subjective, and some
important findings may be missing.  I apologize for the omissions.
Fortunately, there will be many presentations at this conference by the
representatives of most groups.  Also, many important results may be found
at the IAP conference: ``Variable Stars and the Astrophysical Returns of the
Microlensing Searches'' in Paris in 1996 (cf. Ferlet et al. 1997).
The likely prospects for the future of large scale automated surveys is
also described in this paper.

\section{Some Results Related to Pulsating Stars}

There have been many papers about pulsating stars by the EROS and MACHO teams 
(Alcock et al. 1995, 1996, 1997a,b, 1998, 1999a, Bauer et al. 1999,
Beaulieu et al. 1997a,b, Sasselov et al. 1997).  The single most spectacular
result was the period --luminosity ($P-L$) diagram for pulsating stars in the
Large Magellanic Cloud by
the MACHO collaboration.  For the first time the two sequences of
Cepheids: those pulsating in the fundamental mode and those pulsating in
the first overtone were very clearly separated.
In addition, several distinct $P-L$ sequences for red 
giants were also seen for the first time in the MACHO
diagram.  I trust it will be presented later at this conference.

Another spectacular result was obtained by the DUO collaboration, at the
time a single graduate student, Christophe Alard.  Using several hundred
Schmidt camera plates from ESO covering a $5^{\circ} \times 5^{\circ}$ square
in the sky near the galactic bulge, Alard (1996) discovered $ \sim 15,000 $
periodic variables and selected $ \sim 1,500 $ RRab stars.  Using 
two photometric bands he constructed a reddening-independent histogram
of their distance moduli.  Two peaks appeared in the distribution: one
corresponding to the distance of $ \sim 8 $ kpc, the second to the distance
of $ \sim 24 $ kpc, i.e. to the galactic bulge and to the Sagittarius dwarf
galaxy, respectively.  The RR Lyrae variables turned out to be excellent
tracers of the extent of that recently discovered galaxy, extending its size
well beyond the initial estimate (Ibata et al. 1995).

Somewhat less spectacular, but very useful, was the photometry of
over 200 RR Lyrae variables in the Sculptor dwarf galaxy by the OGLE
collaboration (Kaluzny et al. 1995).  This was promptly used by Kov\'acs
\& Jurcsik (1996) to establish a very good correlation between the shape
of the RR Lyrae light curves and the absolute magnitudes.

The recent series of OGLE papers on Cepheids in the Magellanic Clouds (Udalski
et al. 1999a,b,c,d) made a catalog of $ \sim 1,300 $ Cepheids in the LMC, and
a total of $ \sim 340,000 $ photometric measurements in standard $I$, $V$, and 
$B$-bands, accessible over the Internet (Udalski et al. 1999d).
Note, that just several months ago a paper was posted on astro-ph
(Lanoix et al. 1999) presenting ``an exhaustive compilation of all
published data of extragalactic Cepheids''.  The total number of
measurements in the compilation was about 3,000, i.e. two orders
of magnitude fewer than in the paper by Udalski et al. (1999d).

\begin{figure}
\plotfiddle{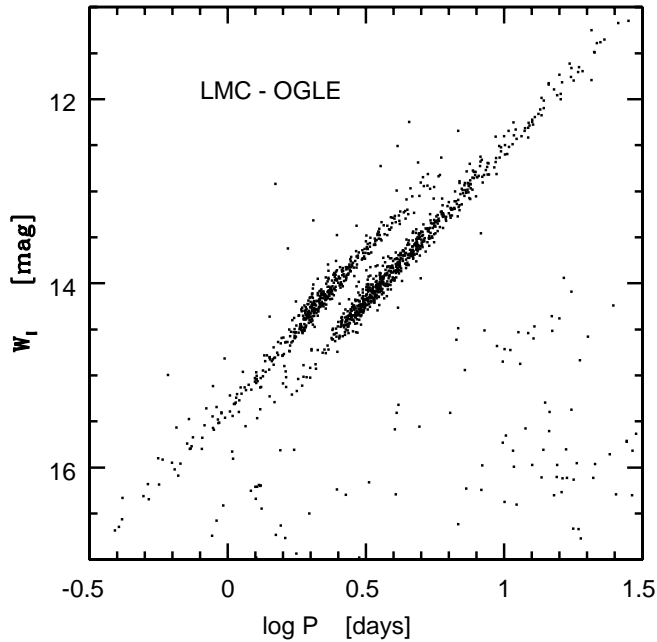}{8cm}{0}{70}{70}{-210}{-110}
\caption{Period - luminosity relation for the LMC Cepheids based on public
domain OGLE data (Udalski et al. 1999d).  The lower and the upper sequences
are populated by Cepheids pulsating in the fundamental mode and in the first
overtone, respectively.  The reddening independent quantity $ W_I $ is 
defined with the eq. (1).}
\end{figure}

The $P-L$ relation for the LMC Cepheids based on the OGLE
public domain data is shown in Fig. 1.
The magnitude $ W_I $ is the interstellar reddening-independent combination of $I$-band and $V$-band photometry:
$$
W_I \equiv I - 1.55 (V-I) .
\eqno(1)
$$
All these refer to the intensity-averaged mean magnitudes.

For each Cepheid a difference $ \Delta W_{O-C} $ can be calculated
between the actual value of $ W_I $ and the line fitted to the fundamental
mode pulsators, given by Udalski et al. (1999c, Table 1) as 
$$
W_{I,C} = -3.277 \log P + 15.815 , \hskip 1.0cm
\Delta W_{O-C} \equiv W_I - W_{I,C} ,
\eqno(2)
$$
where $ P $ is the Cepheid period, in days.
A histogram of the $ \Delta W_I $ values is shown in Fig. 2.  A very clear
separation of the fundamental mode and the first overtone mode pulsators
is apparent in both Figures.

\section{Prospects for the Near Future}

In my view the single most important outcome of the microlensing searches
is the practical demonstration that it is feasible to monitor tens of
millions of stars per night, making up to $ \sim 10^{10} $ photometric 
measurements per year, using very modest instrumentation: a 1-m class 
telescope, a CCD camera, and several PC-class computers.  There can be
no doubt that this technology will develop into much larger surveys.  

\begin{figure}
\plotfiddle{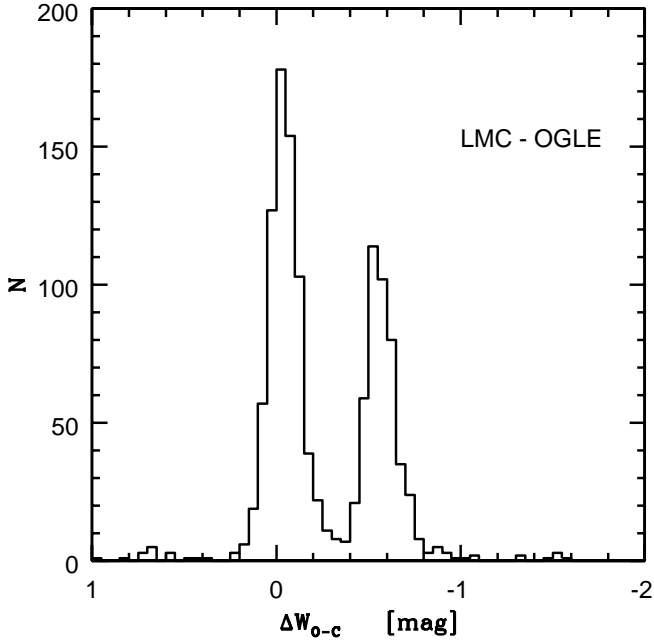}{8cm}{0}{70}{70}{-220}{-150}
\caption{The luminosity function of the LMC Cepheids is shown with respect
to the linear $P-L$ relation given with the eq. (2).  The 
two peaks correspond to the fundamental mode (left) and the first overtone
(right) pulsators.  The definition of the
magnitude difference $ \Delta W_{O-C} $ is given with the eq. (2).}
\end{figure}

The fates of the current projects vary from case to case.  The largest
of them, MACHO, will suffer the ultimate Y2K problem: it will
be terminated on December 31, 1999.  As far as I know, EROS will continue
till at least 2002, while there is no time limit for OGLE.  Currently
OGLE uses a single 2K $\times$ 2K CCD camera in a drift scan mode (cf. Udalski
et al. 1997, OGLE-II) to monitor $ \sim 30 \times 10^6 $ stars per night.
A new mosaic 8K $\times$ 8K CCD camera will become operational in the
year 2000, and it will be used in a still-frame mode.  The data rate will
increase by a factor $ \sim 10 $, leading to the detection of $ \sim 500 $
microlensing events per year with the OGLE-III system.

While hardware upgrades are important, the same is true about software.
The first attempt to use the image subtraction technique to search for 
gravitational microlensing was published by Tomaney \& Crotts (1996).
Preliminary application of image subtraction by the MACHO
collaboration increased the number of detected microlensing events
by a factor $ \sim 2 $ (Alcock et al. 1999b,c).  

New, very powerful image subtraction software has been recently developed
by Alard \& Lupton (1997) using OGLE data.  It has been applied to the old
OGLE microlensing data, dramatically improving photometric accuracy (Alard 
1999a).  Olech et al. (1999) used it to detect RR Lyrae variables all the
way to the center of a globular cluster M5 with ground-based CCD images.
Wo\'zniak et al. (1999) developed with it a real-time photometry system for
Huchra's lens (2237+0305).  A full data pipeline based on Alard \& Lupton
(1997) and Alard (1999b) software is currently under development 
by Wo\'zniak (1999).  The goal is to analyze all OGLE-II
galactic bulge data.  It is likely that incorporation of this
software in the forthcoming OGLE-III system will result in a real-time
detection rate of $ \sim 1,000 $ microlensing events per year.

\begin{figure}
\plotfiddle{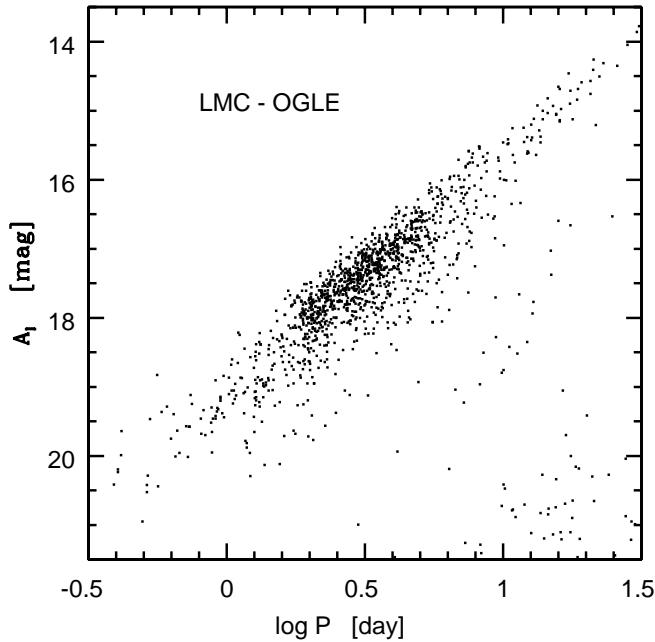}{8cm}{0}{70}{70}{-210}{-110}
\caption{Period - flux amplitude relation for the LMC Cepheids based on public
domain OGLE data (Udalski et al. 1999d).  The flux magnitude in the I-band,
$ A_I $, is defined as the magnitude equivalent of the first term in the 
Fourier expansion; it is not a full amplitude.}
\end{figure}

Image subtraction generates blending-independent determination of stellar
flux variations.  This naturally leads to a period -- flux amplitude 
($P-A$) rather than to $P-L$ relation for Cepheids.  Let
the flux of radiation from a star and its full flux amplitude be defined for
the $I$-band as
$$
F_I \equiv 10^{-0.4 I} , \hskip 0.7cm
\Delta F_I = F_{I,max} - F_{I,min} , \hskip 0.7cm
\Delta m_{F,I} \equiv -2.5 \log _{10} \left( \Delta F_I \right) .
\eqno(3)
$$
Note, that $ \Delta F_I $ is the quantity measured using
the image subtraction method, and $ \Delta m_{F,I} $ is the same quantity
expressed in magnitudes, which are the units preferred by astronomers.
Also note, that both quantities: $ \Delta F_I $ and $ \Delta m_{F,I} $
are independent of blending, and can be reliably measured in the most
crowded environments, without any bias.  Unfortunately, the period - flux
amplitude relation has significantly more scatter than the period - 
luminosity relation, as it is apparent in Fig. 3.

The blending of a Cepheid with images of nearby bright stars
makes it appear brighter than it really is.
Recently, Mochejska et al. (1999) and Stanek \& Udalski (1999)
pointed out that this may lead to a systematic underestimate of distance
moduli of the most distant galaxies studied by the HST Key Project. 
We expect that a $P-A$ relation will eliminate this systematic error.  
However, it is not known if the $P-A$ relation is as universal as the
$P-L$ relation appears to be.

The microlensing searches do not have a monopoly for generating huge sets
of photometric measurements.  DIRECT is a project searching for pulsating
and eclipsing variables in nearby galaxies M31 and M33 (Kaluzny et al. 1998, 
1999, Stanek et al. 1999).  A large number of Cepheids discovered by this
project will provide the first opportunity to verify the $P-A$
relation when combined with the OGLE results for the Magellanic Clouds.

Several large projects using very small instruments are searching
for optical flashes which may accompany gamma-ray bursts (ROTSE: Akerlof
et al. 1999; LOTIS: Williams et al. 1999).  They cover all sky every night
down to $ \sim 14 $ mag.  The first spectacular result was obtained by the
ROTSE collaboration on 1999 January 23: the detection of an optical flash
brighter than 9 mag from the burst GRB 990123, which was at the redshift 
$ z = 1.6 $ (Akerlof et al. 1999).  The archive of ROTSE is
a gold mines for studies of many kinds of variable objects, including
pulsating stars.  In the 2000 or so square degrees analyzed so far
by McKey and his collaborators $ \sim 2000 $ variables were found, 
$ \sim 1800 $ of them new (McKey 1999).

Another project, ASAS, is in its early stages of development (Pojma\'nski
1997, 1998), but with the intent to cover the whole sky with nightly 
observations of all objects brighter than $ \sim 14 $ mag initially, and to
go deeper with time.  This project has real-time photometry implemented
from the start, and it has already demonstrated that even among stars as 
bright as $ 11 - 12 $ mag, over 70\% of all variables have not been 
cataloged, and they are waiting to be discovered.

\section{Prospects for the Distant Future}

There are many ideas about future expansion of sky monitoring, and many
projects at various stages of implementation, planning or dreaming.
The most ambitious of these is the idea of the Dark Matter Telescope
(DMT), as proposed by Angel et al. (1999) and by Armandroff et al. (1999).
This dream is to built an 8.4-m telescope with a 3$^{\circ}$ field imaged at 
f/1.25, with the focal surface filled with $ 1.4 \times 10^9 $ pixels in
a very large mosaic CCD camera.  The primary science objective
is the determination of the distribution of dark matter in the
universe by measuring weak gravitational lensing of galaxies over a
large part of the sky.  However, such an instrument could perform other
spectacular tasks, as it could reach $ V = 24 $ mag in a 20-s exposure,
and cover the whole sky to this depth in about 4 clear nights.  While the
hardware sounds very impressive, the software capable of handling in a
meaningful way such a data rate may be even more challenging.  Data
access will not be easy either.

The DMT is an undertaking on such a huge scale that
a research team even larger than the MACHO or EROS
teams will be needed, first to obtain the necessary funds, next to implement
the project, and finally to get science out of it.  Does this imply that
there will be no room for small teams in the astronomical future?  I do
not think so.  With all its power a DMT-like instrument will be useless
for the discovery of optical flashes like the one associated with the
GRB 990123 (Akerlof et al. 1999).  In order to detect such flashes
independently of gamma-ray burst triggers a very different and far less
expensive instrument is needed to continuously monitor
all sky down to 14 mag, or so.  Note that ROTSE and LOTUS 
cover $ \sim 400 $ square degrees in a single exposure with
four telephoto lenses, each equipped with a 2K $\times$ 2K CCD camera.  To cover
$ \sim 10,000 $ square degrees
we need $ \sim $ 25 small systems like ROTSE or LOTIS.  The
likely cost of so many instruments would be in a relatively modest range
$ 1 - 2 \times 10^6$ dollars.  This level of funding is within reach of
a small team.  For example, the total hardware cost of the OGLE was
$ \sim 1.3 \times 10^6 $ dollars.

The large membership of the MACHO and EROS teams created a misleading
impression that this is the only way for the future.  However, the
current data rate
of OGLE is within a factor 2 or so of the current data rate of MACHO,
yet the OGLE team has only 7 members, most of them students.  Next year the
OGLE data rate will increase by a factor $ \sim 10 $, with one or two 
more students joining the project.  ASAS has only one senior person
and one part-time student, yet within a year its data rate is likely to equal
the current data rate from either ROTSE and LOTIS, and all data will be 
processed in real time.  The most spectacular example of a capability of a 
small team has been demonstrated by DUO, with nearly all data analysis and
writing of science papers done by just one graduate student, Christophe Alard.

Obviously, it is not possible for a small team to make
full use of all the data.  Hence, OGLE (and ASAS in future) makes its data
public domain as soon as feasible.  The bottleneck is quality control.
However, once the data pipeline is fully debugged, it might be possible for
the software to handle quality control nearly in real time.  It might be possible
to make most data public domain right away.  Would this pose a threat to the
scientific recognition of a small team?  I do not think so.  Whoever
uses the data will have to give full credit to the source of the data.
Also, members of a small team could prepare new software needed
to take full advantage of any new hardware upgrades ahead of time, and
this way have a head start over the competition.  It is conceivable that 
the scientific recognition of a team making good data public domain could 
be strongly enhanced by the scientific results obtained with their data by
others.

Several bottlenecks have to be overcome.  The most difficult
and expensive technical bottleneck is software, as has been learned by
large projects like the Sloan Digital Sky Survey (SDSS) and the 2MASS.
It has not been demonstrated yet that a truly large data set can be 
made public domain at a reasonable cost.
Perhaps even more difficult are psychological and sociological problems:
will the authors of useful public domain data get enough
recognition to be offered tenured positions at major universities?
I think these are very important issues.  Most of us enjoy working in small
groups, and find it much less pleasant and less efficient to work almost
anonymously in an industrial size mega-team.  Also, managing large teams 
is very complicated and tedious, with huge overheads in time, funds, and
loss of satisfaction.  I think that whenever a project can be meaningfully
broken down into a number of smaller parts it should be divided into
such parts.  It is much better to reference the work of separate but
collaborating groups, rather than to have a paper with dozens of co-authors,
with no clear division of responsibility and credit for various parts
of the work.  Time will tell if small or large teams will lead in scientific
discoveries in the future.

\acknowledgments
I am very grateful to Dr. A. Udalski for providing the data on which Fig. 3
is based.  It is a pleasure to acknowledge the support by NSF grants
AST-9530478 and AST-9820314.


\begin{references}

\reference{} Akerlof, C., Balsano, R. \& Barthelmy, S. et al. (ROTSE) 1999,
Nature, 398, 400

\reference{} Alard, C. (DUO) 1996, \aap, 458, L17 

\reference{} Alard, C. (OGLE) 1999a, \aap, 343, 10   

\reference{} Alard, C. (OGLE) 1999b, astro-ph/9903111

\reference{} Alard, C., Guibert, J. (DUO) 1997, \aap, 326, 1 

\reference{} Alard, C. \& Lupton, R. H. (OGLE) 1997, \apj, 503, 325 

\reference{} Alard, C., Mao, S. \& Guibert, J. (DUO) 1995, \aap, 300, L17 

\reference{} Alcock, C., Akerlof, C. W., Allsman, R. A. et al. (MACHO) 1993, 
Nature, 365, 621

\reference{} Alcock, C., Allsman, R. A., Axelrod, T. S.
et al. (MACHO) 1995, \aj, 109, 1653 

\reference{} Alcock, C., Allsman, R. A., Axelrod, T. S.
et al. (MACHO) 1996, \aj, 111, 1146 

\reference{} Alcock, C., Allsman, R. A., Alves, D.
et al. (MACHO) 1997a, \apj, 474, 217 

\reference{} Alcock, C., Allsman, R. A., Alves, D.
et al. (MACHO) 1997b, \apj, 482, 89 

\reference{} Alcock, C., Allsman, R. A., Alves, D.
et al. (MACHO) 1998, \aj, 115, 1921 

\reference{} Alcock, C., Allsman, R. A., Alves, D.
et al. (MACHO) 1999a, \apj, 511, 185 

\reference{} Alcock C., Allsman, R. A., Alves, D.
et al. (MACHO) 1999b, \apj, 521, 602

\reference{} Alcock C., Allsman, R. A., Alves, D.
et al. (MACHO) 1999c, \apjs,  124, 171

\reference{} Angel, R., Lesser, M., Sarlot, R., \& Dunham, T. (DMT) (1999) 

\reference{} Armandroff, T., Bernstein, G., Dell'Antonio, I. et al. (DMT) (1999)

\reference{} Aubourg, E., Bareyre, P., Brehin, S. et al. (EROS) 1993, Nature,
365, 623

\reference{} Bauer, F., Afonso, C., Albert, J. N. et al. (EROS) 1999, \aap, 
348, 175 

\reference{} Beaulieu, J. P., Krockenberger, M., Sasselov, D. D. et al. (EROS)
1997a, \aap, 318, L47 

\reference{} Beaulieu, J. P., Sasselov, D. D., Renault, C. et al. (EROS) 1997b,
\aap, 321, L5 

\reference{} Ferlet, R., Maillard, J.-P. \& Raban, B. (Editors) 1997, Variable
Stars and the Astrophysical Returns of the Microlensing Searches, Cedex, France,
Editions Frontieres

\reference{} Ibata, R. A., Gilmore, G. \& Irwin, M. J. 1995, \mnras, 277, 781

\reference{} Kaluzny, J., Kubiak, M., Szyma\'nski, M. et al. (OGLE) 1995,
\aaps, 112, 407 

\reference{} Kaluzny, J., Stanek, K. Z., Krockenberger, M. et al. (DIRECT) 
1998, \aj, 115, 1016

\reference{} Kaluzny, J., Mochejska, B. J., Stanek, K. Z. et al. (DIRECT) 1999,
\aj, 118, 346

\reference{} Kov\'acs, G. \& Jurcsik, J. 1996, \apj, 466, L17 

\reference{} Lanoix, P., Garnier, R., Paturel, G. et al. 1999, astro-ph/9904027

\reference{} McKay, T. 1999, private communication

\reference{} Mochejska, B. J., Macri, L. M., Sasselov, D. D., \& Stanek, K. Z.
(DIRECT) 1999, astro-ph/9908293

\reference{} Olech, A., Wo\'zniak, P. R., Alard, C. et al. 1999,
astro-ph/9905065

\reference{} Pojma\'nski, G. (ASAS) 1997, Acta Astronomica, 47, 467

\reference{} Pojma\'nski, G. (ASAS) 1998, Acta Astronomica, 48, 35

\reference{} Sasselov, D. D., Beaulieu, J. P., Renault, C. et al. (EROS) 1997,
\aap, 324, 471 

\reference{} Stanek, K. Z. \& Udalski, A. (OGLE) 1999, astro-ph/9909346

\reference{} Stanek, K. Z., Kaluzny, J., Krockenberger, M. et al. (DIRECT) 
1999, \aj, 117, 2810

\reference{} Tomaney, A. B., \& Crotts, A. P. S. 1996, \aj, 112, 2872

\reference{} Udalski, A., Szyma\'nski, M., Kaluzny, J. et al. (OGLE) 1993,
Acta Astronomica, 43, 289 

\reference{} Udalski, A., Szyma\'nski, M. \& Kubiak, M. (OGLE) 1997, Acta
Astronomica, 47, 319 

\reference{} Udalski, A., Soszy\'nski, I., Szyma\'nski, M. et al. (OGLE) 1999a,
Acta Astronomica, 49, 1  

\reference{} Udalski, A., Soszy\'nski, I., Szyma\'nski, M. et al. (OGLE) 1999b,
Acta Astronomica, 49, 45  

\reference{} Udalski, A., Szyma\'nski, M., Kubiak, M. et al. (OGLE) 1999c, 
Acta Astronomica, 49, 201 

\reference{} Udalski, A., Soszy\'nski, I., Szyma\'nski, M. et al. (OGLE) 1999d,
Acta Astronomica, 49, 223  

\reference{} Williams, G. G., Park, H. S., Ables, R.  et al. (LOTIS) 1999, 
\apj, 519, L25

\reference{} Wo\'zniak, P. (OGLE) 1999, in preparation

\reference{} Wo\'zniak, P., Alard, C., Udalski, A. et al. (OGLE) 1999, 
astro-ph/9904329

\end{references}
\end{document}